\documentclass[10pt,twocolumn]{article}
\usepackage{latexsym}
\usepackage{amsmath,url}
\usepackage[colorlinks=true,citecolor=black,linkcolor=black]{hyperref}
\usepackage{amstext,amsfonts,epsf,epsfig,pifont}
\usepackage{graphicx,bm,amsmath,natbib}
\usepackage[normalmargins,normalindent,normalbib]{savetrees}
\usepackage[hmargin=.8in,vmargin=0.8in]{geometry}

\begin{document}

\title{{\normalsize Submitted for publication in {\em Geophysical Research Letters}}\\Extracting waves and vortices from Lagrangian trajectories}
\author{J. M. Lilly\footnote{NorthWest Research Associates, Redmond, Washington, USA.}, R. K. Scott\footnote{Department of Applied Mathematics, University of St Andrews, St Andrews, Scotland, UK.}, and S. C. Olhede\footnote{Department of Statistical Science, University College London, London, England, UK.}}

\maketitle

\begin{abstract}
A method for extracting time-varying oscillatory motions from time series records is applied to Lagrangian trajectories from a numerical model of eddies generated by an unstable equivalent barotropic jet on a beta plane.   An oscillation in a Lagrangian trajectory is represented mathematically as the signal traced out as a particle orbits a {\em time-varying} ellipse, a model which captures wavelike motions as well as the displacement signal of a particle trapped in an evolving vortex.  Such oscillatory features can be separated from the turbulent background flow through an analysis founded upon a complex-valued wavelet transform of the trajectory.  Application of the method to a set of one hundred modeled trajectories shows that the oscillatory motions of Lagrangian particles orbiting vortex cores appear to be extracted very well by the method,  which depends upon only a handful of free parameters and which requires no operator intervention.  Furthermore, vortex motions are clearly distinguished from wavelike meandering of the jet---the former are high frequency, nearly circular signals, while the latter are linear in polarization and at much lower frequencies.  This suggests that the proposed method can be useful for identifying and studying vortex and wave properties in large Lagrangian datasets.  In particular, the {\em eccentricity} of the oscillatory displacement signals, a quantity which is not normally considered in Lagrangian studies, emerges as an informative diagnostic for characterizing qualitatively different types of motion.
\end{abstract}

\section{Introduction}
\nocite{lilly09-asilomar,lilly12-itsp,lilly09-itsp,lilly12-itsp,lilly06-npg}

Understanding the role of long-lived eddies in the global ocean circulation is a major topic in oceanographic research.  Lagrangian floats and drifters constitute an invaluable  platform for observing the relatively small spatial and temporal scales associated with such vortices, and indeed many dozens of publications have examined vortex properties from regional Lagrangian experiments,  e.g. \citet{armi89-jpo}, \citep{flament01-jfm}, and \citep{shoosmith05-dsr}.   Systematic investigation of the now-extensive historical set of Lagrangian drifter  and float trajectories is, however, hampered by a technical limitation---the lack of a reliable and precise method to meaningfully separate vortex motions from the background flow.  Various approaches have been proposed \citep[e.g.][]{armi89-jpo,flament01-jfm,lankhorst06-jaot}.  However, the fact that this problem remains unsolved is evidenced by the fact that large-scale studies either continue to rely on traditional subjective identification \citep[e.g.][]{shoosmith05-dsr}, or else to focus on measures of the effect of vortices on trajectories rather than the properties of the vortices themselves \citep{griffa08-grl}.

Recently a new and general method has been developed \citep{lilly06-npg,lilly09-asilomar,lilly12-itsp}, grounded in nonstationary time series theory, which permits the automated identification, extraction, and analysis of time-varying oscillatory features of unknown frequency---such as the signature of a particle trapped in a vortex or advected by a wave.  Here the method is applied to Lagrangian trajectories from a numerical simulation in order to illustrate the possibility of accurately extracting and distinguishing vortex currents and wavelike motions.  All relevant analysis software is freely distributed to the community as a part of a Matlab\textregistered~toolbox, available at \url{http://www.jmlilly.net}.

\section{Numerical Model}
For an idealized numerical model generating eddies as well as background variability, we choose an equivalent barotropic quasigeostrophic model of an initially unstable jet on a  beta plane.  The model integrates the equation for conservation of potential vorticity following the geostrophic flow
\begin{equation}\label{vorticityequation}
\left(\frac{\partial}{\partial t}+   {\bf k}\cdot\nabla\Phi \times \nabla \right)\left(\nabla^2\Phi - \Phi/L_D^2 + \beta y \right)=0
\end{equation}
where $\Phi$ is the streamfunction, $\beta\equiv df / d\vartheta$ is the derivative of the Coriolis frequency $f$ with latitude $\vartheta$, the parameter $L_D$ is the Rossby radius of deformation, and $\mathbf{k}$ is the vertical unit vector.  The model is initialized at time $t=0$ with an eastward jet of strength $U$ and width $Y$ having a profile, with $\mathbf{i}$ being the eastward unit vector, given by
\begin{equation}
\left.\mathbf{u}\right|_{t=0}=\mathbf{i} \left\{\begin{array}{lc}U\cos^2\left(\frac{y}{Y}\frac{\pi}{2}\right)&\quad |y|/Y\le 1\\
0 & \quad  |y|/Y >1 \end{array}\right.
\end{equation}
which corresponds to a maximum initial vorticity anomaly within the jet of $\zeta_o\equiv \pi U/(2Y)$.

Parameters are chosen to give a strong jet with a deformation radius that is small compared to the radius of the earth.   The central latitude ($y=0$) at the jet axis is set to $\vartheta=45^\circ$~N, the jet width $2Y$ and deformation radius $L_D$ are both 80~km, and the maximum initial velocity is 2.08 m~s$^{-1}$. These choices give a jet Rossby number $Ro\equiv\zeta_o/f=0.80$, and a value of $\beta L_D/\zeta_o$ of $1/(20\pi)\approx 1/60$---meaning that the jet relative vorticity anomaly is much larger than the change in planetary vorticity over one deformation radius, or over the jet width.  This system is a convenient way of generating eddies,  and is not intended to represent a particular oceanic current.   The initial condition is let to freely evolve for 360 days with a time step of $5\times10^{-5}\times360$ days $\approx26$ minutes in a domain of length $2\pi\times 400 \approx 2500$~km on each side.  The model is seeded with 100 drifters that all initially lie along the $y=0$ line, sampled every 10 time steps or 4.3 hours.

A snapshot of the model is shown in Fig.~1, together with overlays of ellipses characterizing oscillatory Lagrangian variability from our subsequent analysis.  Vortices are generated by barotropic instability and then tend to drift westward as well as meridionally due to nonlinear beta drift \citep[e.g.][]{lam01-jfm}, with anticyclones propagating equatorward and cyclones propagating poleward.  Dipole interactions are also seen, as captured by the red/blue pair of ellipses in the upper left quadrant of Fig.~1, although these tend to eventually break down.  An informative animation of snapshots such as the one shown in Fig.~1 over the entire model run duration, but with the ellipses color coded by drifter number for visual clarity, may be found in the file \texttt{vortexmovie.mov} at \url{ftp://ftp.nwra.com/outgoing/lilly/vortex}.  Trajectories are shown in Fig.~2a, with the dominating presence of vortex motions apparent as tightly looping or cycloidal curves reaching northwestward and southwestward. 

\begin{figure}[t]
\centering
 \includegraphics[width=20pc]{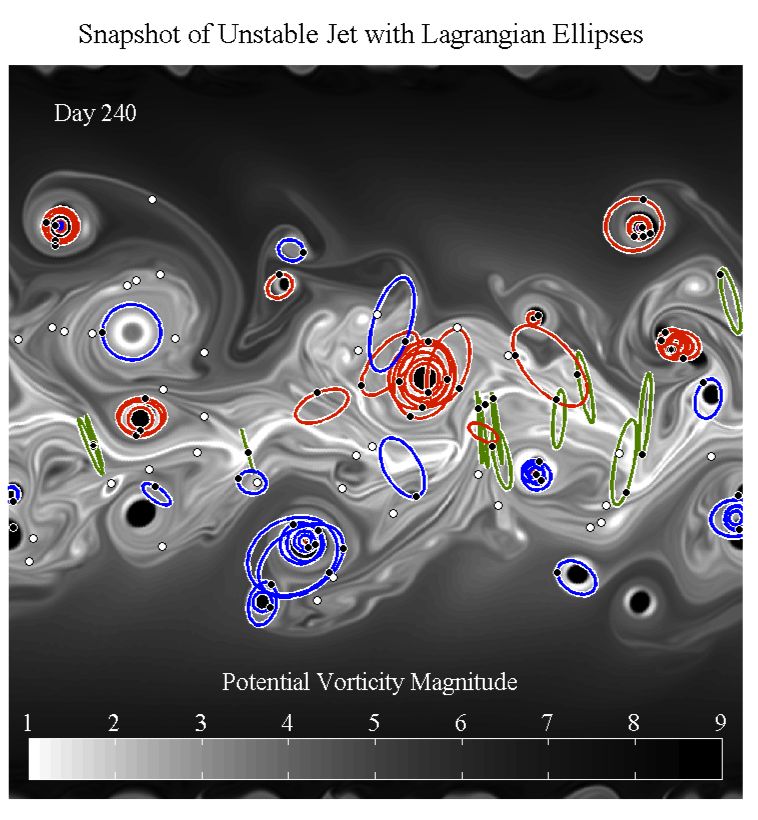}
\caption{A snapshot of an unstable eastward equivalent barotropic jet on a mid-latitude beta plane at day 240, with domain details as discussed in the text. The shading is the absolute value of the quasigeostrophic potential vorticity appearing in (\ref{vorticityequation}).  Estimated instantaneous ellipses due to Lagrangian oscillatory motions, created as in Section~3, are overlaid.  Highly eccentric ellipses with eccentricity $\varepsilon>0.95$ are shown in green, while positively-rotating and negatively-rotating ellipses with $\varepsilon\le0.95$ are shown in red and blue, respectively.  Dots mark the instantaneous locations of 100 Lagrangian particles initially deployed along the jet axis. Black dots indicate particles in which an oscillatory signal is detected at this moment, while white dots indicate the locations of other particles. }\label{snapshot}
\end{figure}

\section{Analysis Method}

A {\em modulated elliptical signal}, introduced by \citet{lilly10-itsp}, is a time-varying oscillation in two dimensions.  Such a signal is expressed in matrix form as
\begin{equation}
 \widetilde{\mathbf{x}}(t) =
  \begin{bmatrix}\cos\theta(t)&-\sin\theta(t) \\ \sin\theta(t)& \cos\theta(t) \end{bmatrix}
  \begin{bmatrix}a(t)  \cos \phi(t) \\ b(t) \sin\phi(t)\end{bmatrix}\label{signalmodel}
\end{equation}
which is the parametric equation for a particle orbiting a time-varying ellipse with semi-major and semi-minor axes $a(t)$ and $|b(t)|$, where $a(t)>|b(t)|>0$, and with its major axis inclined at an angle $\theta(t)$ with respect to the $x$-axis.  The phase $\phi(t)$ gives the instantaneous position of the particle along the periphery of ellipse.  The particle orbits the ellipse in the mathematically positive or negative direction according to the sign of $b(t)$.

Details of the modulated elliptical signal, including conditions for associating a {\em unique} set of time-varying ellipse parameters to a given oscillatory signal $\widetilde{\mathbf{x}}(t)$, are discussed in \citet{lilly10-itsp}.  An important special case is that of a familiar pure sinusoidal oscillation in two dimensions; however, the model (\ref{signalmodel}) is considerably more general.  A practical constraint is that the ellipse properties should vary slowly compared with the timescale $2\pi/\frac{d}{dt}\phi(t)$ over which the particle orbits the ellipse, in order that the subsequent analysis method have small errors \citep{lilly12-itsp}.

The modulated elliptical signal with slowly-varying ellipse parameters is a good model for the displacement signal of a particle trapped in an evolving vortex.  This class of signals includes Lagrangian displacements due to steady circular vortex solutions, steadily strained or sheared elliptical anticyclones \citep{ruddick87-jpo}, and low-frequency periodic oscillations of an elliptical shallow water vortex \citep{young86-jfm,holm91-jfm}, all observed with instruments that may be experiencing a drift through the vortex in addition to the vortex currents themselves.

Ellipse size, shape, and frequency are usefully characterized as follows. The ellipse shape is described by the eccentricity $\varepsilon(t)\equiv\sqrt{1-b^2(t)/a^2(t)}$.  The geometric mean radius and geometric mean velocity are defined as
\begin{equation}
R(t)  \equiv  \sqrt{a(t)|b(t)|}, \quad 
V(t)  \equiv \mathrm{sgn}\left\{ b_\mathbf{u}(t)\right\} \sqrt{ a_{\mathbf{u}}(t) |  b_{\mathbf{u}}(t)|}
\end{equation}
where the latter quantity is found by writing the velocity $\widetilde{\mathbf{u}}(t)\equiv\frac{d}{dt}\widetilde{\mathbf{x}}(t)$ as a time-varying ellipse of the form  (\ref{signalmodel}) but with a different set of ellipse parameters, denoted $a_{\mathbf{u}}(t)$, $b_{\mathbf{u}}(t)$, etc.; this may be accomplished algebraically from the parameters of $\widetilde{\mathbf{x}}(t)$, see Appendix~E of \citet{lilly06-npg}.   The ratio of the two quantities $V(t)$ and $R(t)$ defines a frequency $\varpi (t)  \equiv V(t)/R(t)$ which we call the {\em geometric frequency}.

A Lagrangian trajectory can then be represented as the sum of a number $M$ of different oscillatory displacement signals $\widetilde{\mathbf{x}}^{\{m\}}(t)$, each of the form (\ref{signalmodel}), plus a residual:
\begin{equation}\label{signalpartition}
\mathbf{x}(t)=\begin{bmatrix}x(t) \\ y(t) \end{bmatrix}
=  \sum_{m=1}^M \widetilde{\mathbf{x}}^{\{m\}}(t)  +\bm{\epsilon}(t).
\end{equation}
The residual signal $\bm{\epsilon}(t)$ includes the turbulent background flow, as well as any non-oscillatory component such as a mean flow or the systematic self-propagation tendency of a vortex. The oscillatory signals may be of finite duration and may be overlapping in time, so that zero, one, or more than one such signal may be present at each moment.

The problem is then to estimate the  oscillations $\widetilde{\mathbf{x}}^{\{m\}}(t)$ given an observed trajectory $\mathbf{x}(t)$, and from these estimates to characterize the $M$ different ellipse parameters $a^{\{m\}}(t)$, $b^{\{m\}}(t)$, $\theta^{\{m\}}(t)$, $\phi^{\{m\}}(t)$, and so forth.  This is accomplished using a method called ``multivariate wavelet ridge analysis''  \citep{lilly09-asilomar,lilly12-itsp}, leading to estimates $\widehat{\mathbf{x}}_\psi^{\{m\}}(t)$ of the $M$ modulated oscillations in each trajectory; here the subscript ``$\psi$'' indicates the choice of filter or wavelet $\psi(t)$ used in the analysis.  A discussion of the basic idea and implementation of the method, together with details of the parameter settings used here, are provided in a technical appendix.

From the method, we obtain estimates of ellipse properties at each moment.  As an example, observe the good agreement in Fig.~1 between the ellipses, inferred non-locally from individual particles, and the Eulerian structures in the potential vorticity field.

\section{Results}

The results of this analysis are shown in Figs.~2 and~3.  Subtracting the sum over all $M$ estimated oscillations in each time series $\mathbf{x}(t)$ leads to an estimate $\widehat{\bm{\epsilon}}(t)$ of the non-oscillatory background flow, Fig.~2b.   Comparison with the original time series,  Fig.~2a, shows that the tightly looping motions associated with vortices appear to have been nearly completely removed.  What remains behind is observed to consist of curving but disorganized motions, together with systematic motion associated with the jet and with eddy drift.  Even cycloidal features in Fig.~2a, typically low-frequency motion of a particle on the far flank of an eddy, are largely removed.  The ability to perform such a separation on this relative large set of drifters, with no intervention by the analyst, by itself constitutes a technical breakthrough.

\begin{figure*}
\centering
 \noindent\includegraphics[width=37pc]{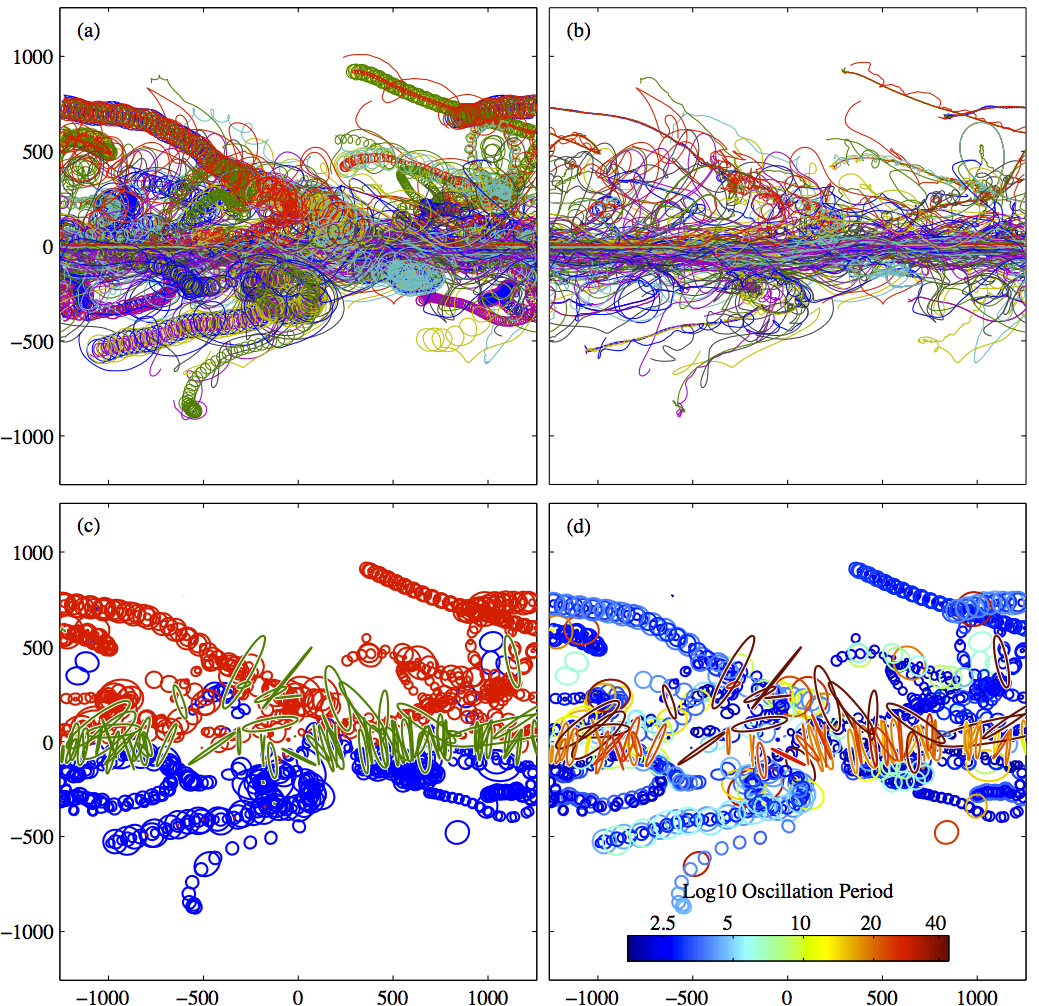}
\caption{Panel (a) shows all Lagrangian trajectories from the unstable barotropic jet simulation, dispersing from their initial location at $y=0$, with each trajectory in a different color.  In (b), the residual curves from the wavelet ridge analysis are presented. Snapshots of ellipses corresponding to the ellipse properties estimated from the wavelet ridge analysis are shown in (c) and (d).  In both panels, highly eccentric ellipses with $\varepsilon>0.95$---typically very low-frequency signals---are plotted with a time step of 1/2 the estimated period, while other ellipses are plotted with the time step of every 3 estimated periods.  The highly eccentric ellipses are plotted in (c) in green, while the remaining anticyclonically rotating ellipses and cyclonically rotating ellipses are plotted in blue and red, respectively.  Finally in (d), the ellipses are colored according to $\log_{10}$ of the estimated geometric period $2\pi/\varpi(t)$, measured in days, as indicated in the color bar.
}\label{vortex-decomposition}  
\end{figure*}

Instantaneous ellipses associated with the estimated oscillatory signals are shown in Fig.~2c,d.   Coloring the ellipses according to their degree of eccentricity and sense of rotation in Fig.~2c reveals nearly exclusively cyclonic motions (red) on the poleward side of the jet and anticyclonic motions (blue) on the equatorward side.   The separation of vortices by their polarity in this system is due to nonlinear beta drift acting on eddies generated within the jet core during the initial adjustment.     
Close inspection reveals some very small circles of opposite color inside some of the eddies; as discussed in the technical appendix, these represent weak, low-frequency signals, likely associated with the presence of exterior opposing vorticity anomalies.  Another type of variability, nearly linear in polarization and oriented meridionally, is observed along the jet axis (green).

A time scale distinction between these two different types of motions can be seen in Fig.~2d, where the color coding represents the instantaneous oscillation period $2\pi/\varpi(t)$ as deduced from the geometric frequency $\varpi(t)$.    The highly eccentric motions along the jet axis are seen to be an order of magnitude lower in frequency than the vortex motions. The former arise as particles in the jet are swept eastward through meanders caused by the deflection of the jet axis by low-frequency Rossby wave variability, as is readily apparent in the animation file \texttt{vortexmovie.mov} at \url{ftp://ftp.nwra.com/outgoing/lilly/vortex} .

A more detailed view of the properties of the estimated elliptical signals is found in the distribution plots on the radius/velocity plane of Fig.~3.  A histogram on the geometric radius $R(t)$ / geometric velocity $V(t)$ plane is formed in Fig.~3a by binning all time points associated with each estimated modulated oscillation in all 100 trajectories.  Note that the slope $V(t)/R(t)$ on this plane is the geometric frequency $\varpi(t)$.  This histogram clearly reveals a Rankine-type vortex profile---a solid-body core plus $1/r$ decay---for cyclonic motions, $V(t)>0$.   As may be expected, the solid-body histogram peak lies along the slope corresponding to the maximum vorticity anomaly of the initial unstable jet profile (gray line); because under the assumption of  solid-body rotation we have $\zeta=2V/R$, this occurs at a frequency of $\zeta_o/2=0.40 f$.

This vortex profile pattern emerges primarily through the superposition of a number of different Lagrangian particles at different locations in different eddies.  On the anticyclonic side the pattern is less evident, since by chance, more particles have ended up in cyclonic eddies compared to anticyclonic eddies in this simulation; this asymmetry is a reminder of the well-known slow convergence of Lagrangian statistics owing to long-term particle trapping  \citep[e.g.][]{pasquero02-prl}.

The low-frequency motions are evident as the symmetric maxima in Fig.~3a  around the horizontal line $V=0$.  The median eccentricity $\varepsilon$ in each bin,  Fig.~3b, shows that the motions in the vortex profile are nearly circular in polarization, while the low-frequency motions are nearly linear, confirming that these two regions on the $R/V$ plane correspond to the two types of features seen in  Fig.~2c,d.  An important point is that the nearly circular vortex motions and the highly eccentric low-frequency motions occupy almost non-overlapping regions of the radius/velocity plane, apart from very large-scale (say $\sim$50~km radius) and low-frequency motions, which may either be generated by jet meander or by circular oscillatory motions on the far flank of a vortex.

\section{Conclusions}

This paper has applied a new analysis method---multivariate wavelet ridge analysis---to the identification of oscillatory motions in Lagrangian trajectories from a numerical simulation of an unstable jet.  It is shown that vortex motions can be effectively extracted from the trajectories and described locally in terms of their time-varying frequency content and ellipse geometry.  Meridional meandering of the jet axis constitutes another strong oscillatory signal in this model, but these motions are clearly distinguished from the vortex motions on account of their much lower frequency, nearly linear shape polarization, and different location in radius/velocity space.  The method is therefore able to unravel the superposition of different processes in individual trajectories, making possible the investigation of these separate processes in isolation from one another.  These results serve as validation supporting the application of this method to large-scale observational studies.  Not addressed here, but left to the future, are a detailed treatment of stochastic errors, the impact of measurement noise, comparison between the Eulerian and Lagrangian perspectives, and a consideration of method performance for other types of motions such as inertial oscillations or baroclinic instability waves.

\begin{figure*}
\centering
\includegraphics[width=37pc]{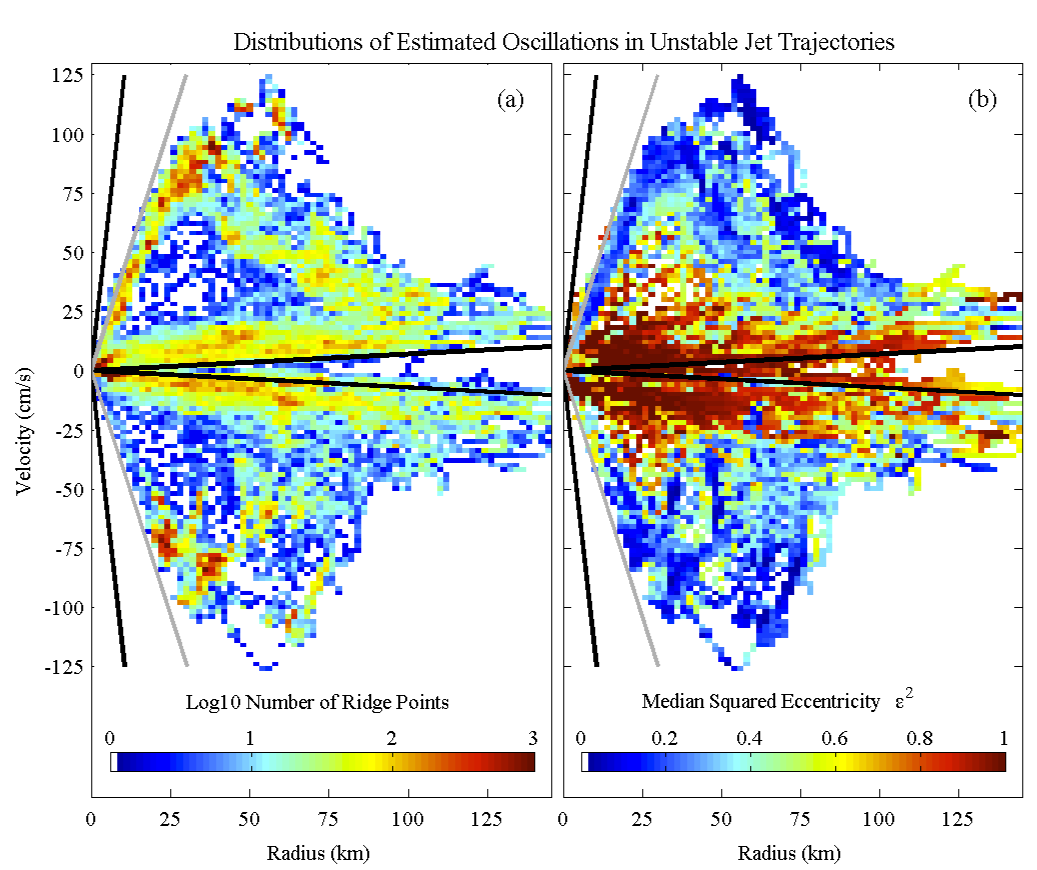}
\caption{Statistics of the estimated modulated elliptical signals shown on the geometric radius/velocity or $R/V$ plane. Positive and negative velocities correspond to cyclonic and anticyclonic motions respectively.  In (a) we see the histogram on the $R/V$ plane occupied by all estimated oscillatory signals from all 100 trajectories, with a logarithmic color axis.  The median eccentricity $\varepsilon$ in each bin is shown in (b).   In both panels, the sloping gray lines correspond to $\pm\zeta_o/2$, the geometric frequency value corresponding (for an eddy in solid-body rotation) to the maximum initial jet relative vorticity anomaly. The black lines are plus or minus the maximum and minimum cutoff frequencies in the analysis.}\label{distributions}
\end{figure*}

\paragraph{Acknowledgements.}  The work of J. M. Lilly was supported  by award \#0751697 from the Physical Oceanography program of the United States National Science Foundation.  The work of S.~C.  Olhede was supported by award \#EP/I005250/1 from the Engineering and Physical Sciences Research Council of the United Kingdom.   The animation file was generated using Image2Movie by Jeffrey J. Early, available at \url{http://jeffreyearly.com/programming/image2movie}.

\appendix
\section{Technical Appendix}

The extraction of individual oscillatory signals from the modeled Lagrangian trajectories is accomplished using a method called ``multivariate wavelet ridge analysis''   \citep{lilly09-asilomar,lilly12-itsp}.  All numerical code associated with this paper is distributed as part of a Matlab\textregistered~toolbox called Jlab, available at \url{http://www.jmlilly.net}.  The file \texttt{makefigs}\_\texttt{vortex} provides the exact processing steps as well as scripts to generate all figures. This appendix describes the basic idea and presents an example. Equation numbers herein refer to the main text.

An {\em analytic wavelet} is a time/frequency localized filter which has vanishing support on negative Fourier frequencies, and which is therefore complex-valued in the time domain, see e.g. \citet{lilly09-itsp} and references therein for details.   The wavelet transform of a real-valued signal vector $\mathbf{x}(t)$ with respect to the wavelet $\psi(t)$ is defined as
\begin{equation*}
\label{vectorwavetrans}
\mathbf{w}_{\mathbf{x},\psi}(t ,s) \equiv  \int_{-\infty}^{\infty} \frac{1}{s}
  \psi^*\left(\frac{\tau-t}{s}\right)\,\mathbf{x}(\tau)\,d \tau
\end{equation*}
and can be seen, on account of the $1/s$ normalization, as a set of bandpass operations indexed by the ``scale''~$s$ which controls the dilation or contraction of the wavelet in time.  The frequency-domain wavelet $\Psi(\omega)$ obtains a maximum magnitude, set to $\Psi(\omega)=2$, at some frequency $\omega_\psi$. This frequency is used to convert scale into a period via $2\pi s/\omega_\psi$.

The particular choice of analyzing wavelet $\psi(t)$ is important in order to minimize errors in the subsequent analysis \citep{lilly09-itsp,lilly12-itsp}.  We use the ``Airy wavelet'' of \citet{lilly09-itsp}, which can be seen as a superior alternative to the popular but problematic Morlet wavelet, as discussed therein.  The Airy wavelet is controlled by a parameter $P_\psi$, with $P_\psi/\pi$ giving the number of oscillations spanning the central window of the wavelet.   We choose $P_\psi/\pi=\sqrt{6}/\pi\approx 0.78$ in order to obtain a high degree of time concentration at the expense of frequency resolution.

{\em Multivariate wavelet ridge analysis} estimates modulated oscillations in a multivariate, or vector-valued, time series by first identifying maxima of the transform magnitude.  A brief introduction to this method may be found in  \citet{lilly09-asilomar}, with further details and bias estimates provided by \cite{lilly12-itsp}.  A {\em ridge point} of $\mathbf{w}_{\mathbf{x},\psi}(t ,s)$ is defined to be a point on the $(t,s)$ or ``time/scale'' plane satisfying
\begin{equation*}
\frac{\partial}{\partial s}\, \left\|\mathbf{w}_{\mathbf{x},\psi}(t ,s)\right\| =  0\label{ampvectorridge},\quad \quad
\frac{\partial^2}{\partial s^ 2}\,  \left\|\mathbf{w}_{\mathbf{x},\psi}(t ,s)\right\| <   0
\end{equation*}
and thus ridge points are locations where the norm of the wavelet transform vector achieves a local maximum with respect to variations in scale. Adjacent ridge points are then connected to each other to yield a single-valued, continuous scale curve as a function of time called a {\em ridge}~$\widehat s(t)$.

A number, say $M$, different ridges $\widehat s^{\{m\}}(t)$ may be present in the same time series, and may overlap in time.   The real part of the wavelet transform along the $m$th ridge may be taken as an estimate of the $m$th oscillatory signal $\widetilde{\mathbf{x}}^{\{m\}}(t)$ in the composite model (5).  That is
\begin{equation*}\label{approximation}
\widehat{\mathbf{x}}_\psi^{\{m\}}(t)\equiv\Re\left\{\mathbf{w}_{\mathbf{x},\psi}\!\left( t, \widehat s^{\{m\}}(t)\right)\right\}= \widetilde{\mathbf{x}}^{\{m\}}(t)+ \Delta\widetilde{\mathbf{x}}_\psi^{\{m\}}(t)
\end{equation*}
where the last term on the right-hand side is an error term.  This may be understood as resulting from the fact that the real part of the wavelet transform at scale $s=\omega_\psi/\omega$  is a bandpass at frequency $\omega$; thus the wavelet ridge estimate combines different pass bands at different times, as is appropriate for the time-varying nature of the signal.  The error term depends upon several factors: (i) the local ratio of the oscillatory signal strength to the background variability strength, (ii) the magnitude of changes in oscillation properties over the wavelet timescale, see the treatment in \citet{lilly12-itsp}, and (iii) the distance between contemporaneous oscillatory signals in frequency compared with the wavelet frequency profile.  A detailed treatment of errors is outside the scope of this paper; instead, we will point to Fig.~2b in the text and the example below as subjective evidence that oscillatory signals have been recovered satisfactorily.

It remains to form estimates of the ellipse parameters in the modulated elliptical signal model (3) associated with each of the $M$ ridges.  Observe that, as there are two parameters on the left-hand side of (3) but four parameters on the right-hand side, the ellipse parameters are underdetermined for a given oscillatory signal $\widetilde{\mathbf{x}}(t)$.    Since the analytic wavelet transform $\mathbf{w}_{\mathbf{x},\psi}\!\left( t, s \right)$ is a complex-valued 2-vector, it consists of four quantities at each time. It has been shown by \citet{lilly06-npg} and \citet{lilly10-itsp} that the complex-valued wavelet transform evaluated along the ridge may be written as
\begin{multline*}
\mathbf{w}_{\mathbf{x},\psi}\!\left( t, \widehat s^{\{m\}}(t)\right) = \\ e^{i\widehat\phi^{\{m\}}(t)}
  \begin{bmatrix}\cos\widehat\theta^{\{m\}}(t)&-\sin\widehat\theta^{\{m\}}(t) \\ \sin\widehat\theta^{\{m\}}(t)& \cos\widehat\theta^{\{m\}}(t) \end{bmatrix}
\begin{bmatrix}\widehat a^{\{m\}}(t) \\  -i \widehat b^{\{m\}}(t) \end{bmatrix}
\end{multline*}
with the four quantities on the right-hand side being the sought-after estimates of the four ellipse parameters at each moment for each ridge. The ``hats'' mark estimated ellipse parameters that are defined implicitly through this equation.

In practical implementation of the wavelet ridge analysis, two numerical thresholds must be introduced.  Firstly a cutoff on the amplitude of ridge points is applied in order to reject very weak oscillations, which we choose here as position oscillations of less than 400 meters (1/100 of the jet half-width $Y$) in magnitude.  Secondly we define the {\em ridge duration}
\begin{equation*}\label{approximation}
D^{\{m\}}\equiv\frac{1}{2\pi} \int_{T^{\{m\}}}\arg \left\{\mathbf{w}_{\mathbf{x},\psi}\!\left( t, \widehat s^{\{m\}}(t)\right)\right\} dt
\end{equation*}
where ``$\arg$'' denotes the complex phase, and the integral is taken over the time interval $T^{\{m\}}$ of the $m$th ridge.  The ridge duration gives the number of complete oscillations executed by the estimated signal;  we reject all ridges that execute fewer than $D^{\{m\}}=2P_\psi/\pi\approx 1.6$ complete cycles, as we find ridges that are too short compared with the wavelet duration are generally spurious.

An example of wavelet ridge analysis is presented in Fig.~A1 for one of the trajectories from the numerical simulation.  The position trajectory $\mathbf{x}(t)$ shown in Fig.~A1a is transformed using the parameter settings discussed above.  104 different scale levels are used which vary logarithmically from a minimum period  $2\pi s/\omega_\psi$ of 0.6 days to a maximum of 103 days.  The norm of the wavelet transform vector is shown in  Fig.~A1b which, like position record itself, has units of kilometers.  In the wavelet transform we see low-frequency variability with periods of $\sim$10~days, characteristic of geostrophic turbulence, and higher-frequency variability with periods of $\sim$1 day reflecting particle motion around vortex cores.  The large spectral gap between them supports our choice of a time-localized wavelet.
\begin{figure*}[t!]
\centering
\includegraphics[width=37pc]{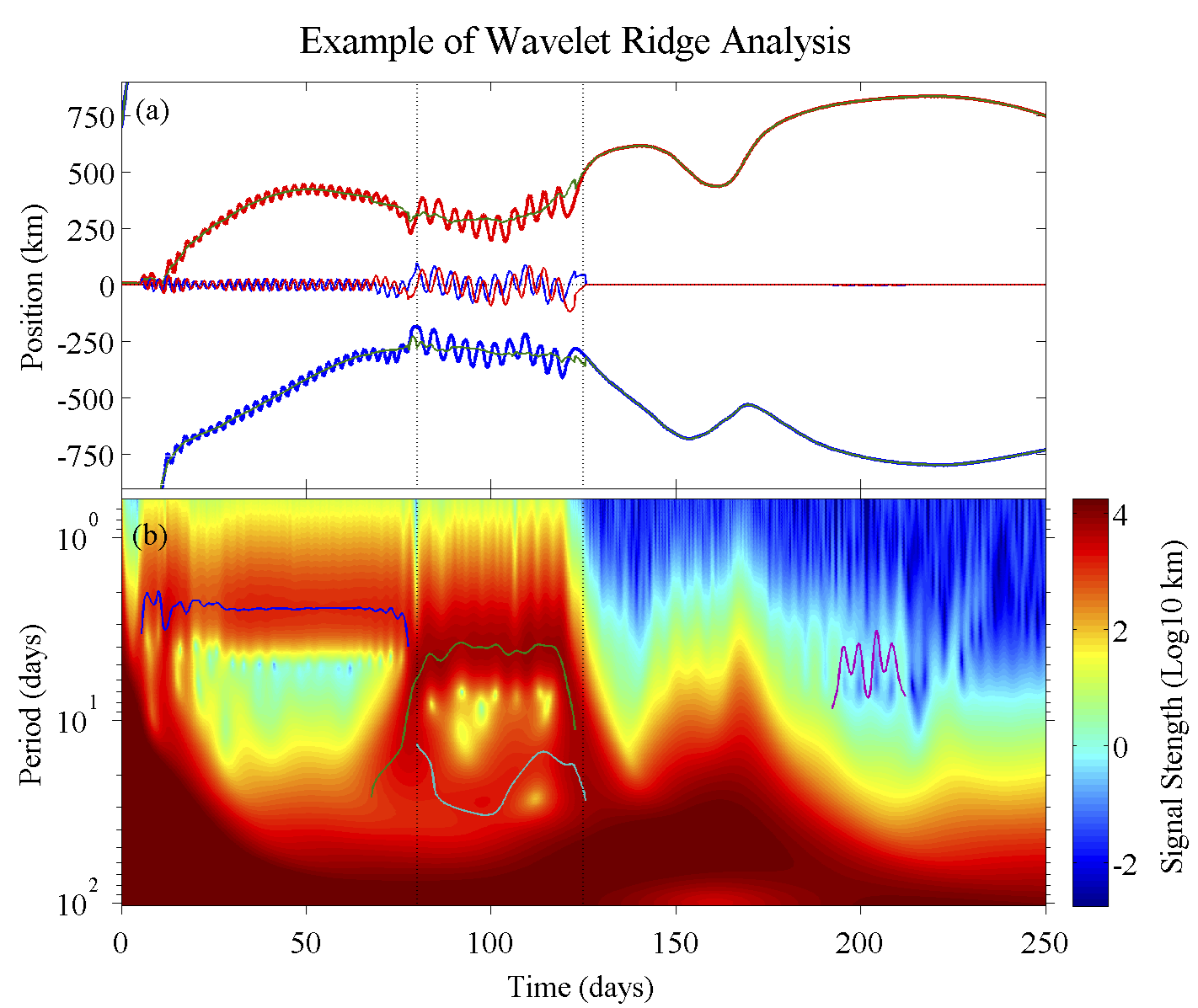}
\caption{An illustration of wavelet ridge analysis. In (a), the east--west and north--south displacement signals associated with a Lagrangian ``particle'' in the model simulation are shown with the blue and red heavy lines, respectively.  Thin blue and red lines show aggregate east--west and north--south oscillatory displacement signals obtained by summing over all estimated modulated oscillations $\widehat{\mathbf{x}}_\psi^{\{m\}}(t)$  at each time.  The green curves are the east--west and north--south residuals $\mathbf{x}(t)-\sum_{m=1}^M\widehat{\mathbf{x}}_\psi^{\{m\}}(t)$.  The wavelet transform norm $\left\|\mathbf{w}_{\mathbf{x},\psi}(t ,s)\right\|$ is shown in (b), with scale converted into period on the $y$-axis.  Four ridge curves are identified and drawn.  The real part of the wavelet transform evaluated along these four curves give the estimated oscillations used in panel (a).  Vertical lines in both panels mark the approximate times of vortex merger events as seen in the animation.}\label{vortex-transform}
\end{figure*}

$M=4$ ridges are detected and are marked in the figure as curves indicating maxima of the wavelet transform with respect to scale.  The real part of the wavelet transform along the $m$th ridge estimates the $m$th oscillatory signal component in the composite model (5), which the analysis is able to isolate from the surrounding variability.  These four estimated signals are then plotted in Fig.~A1a along with the residuals.  It is observed that the oscillatory features in the original trajectory are removed extremely well by the analysis method.

In the central part of the record, two ridges are simultaneously present, with the lower-frequency curve representing an interaction with a distant vortex.   A very weak amplitude ridge, just greater than our cutoff amplitude, is detected near the end of the time series; this is almost certainly spurious and illustrates the reason why we impose an amplitude cutoff; weak ``false positives'' such as this signal do not impact our analysis because their amplitude is so small, and because they appear to occur rarely.  Note that the low-frequency background, while slowly meandering in nature, does not generate ridges because it does not typically result in the appearance of multiple orbits through the same oscillatory structure, and is therefore not characterized as a modulated oscillation by this analysis.

There are two transitions in this time series: one around 80 days, when the main ridge lowers its frequency and a second contemporaneous ridge appears, and a second transition around 125 days, when these two ridges disappear.  It can be seen from watching the animation that these two transitions correspond to two vortex mergers.  In that animation, ellipses associated with this trajectory are drawn in black, and the position of this drifter is indicated with a magenta dot.  During the first merger, the vortex grows in size and its frequency decreases, but there is a remaining distant vortex which causes lower-frequency oscillatory motion that is manifested as the lower of the two ridges.  During the second merger, the particle is ejected in a filament and consequently the oscillatory motions come to an end.  This indicates that transitions in the ridges can have physical meaning and can capture vortex evolution.


\end{document}